# PMU Assisted Power System Parameter Calibration at Jiangsu Electric Power Company

Xiao Lu, *Di Shi, Bin Zhu, *Zhiwei Wang, Jianyu Luo, Dawei Su, Chunlei Xu

State Grid Jiangsu Electric Power Company, Nanjing, Jiangsu, China

*GEIRI North America, Santa Clara, CA, United States

Email: di.shi@geirina.net

*Abstract*—**An online PMU-assisted Power System Parameter Calibration System (PSPCS) was recently developed and implemented at State Grid Jiangsu Electric Power Company (JEPC). PSPCS leverages high-resolution PMU data and data mining techniques to perform online screening of the EMS and Production Management System (PMS) databases for data cleaning, model validation, and parameter calibration. PSPCS calculates transmission line and generator parameters on a regular real-time basis and compares the results with databases to identify record(s) with significant discrepancy, if any. Once consistent discrepancy is observed, the system will raise a flag and further investigation will be initiated, including a novel density-based spatial clustering procedure for parameter/data calibration. A novel metric is proposed to quantify the credibility of PMU-based parameter identification. This paper discusses the proposed methodologies, challenges, as well as implementation issues identified during the development and deployment of PSPCS.**

## I. Introduction

Jiangsu Electric Power Company (JEPC) operates one of the largest provincial power systems in China, with a number of ultra-high voltage DC transmission lines (UHVDC). Hybrid operation of AC and DC, complicated network structure, high penetration of renewables, and great challenges on operation and control are the features of Jiangsu grid. In the development of smart grid, JEPC has spent tremendous efforts in maintaining grid reliability and creating an open infrastructure that can take advantage of advanced technologies and evolving tools. A critical portion of this effort is to construct and maintain an authentic and accurate model of Jiangsu grid.

Phasor measurement unit (PMU) is envisioned to be one of the enabling technologies in smart grid, with many potential applications including parameter/system identification, system monitoring, relay protection, state estimation, wide-area visibility, control applications, etc. At present, JEPC has full PMU coverage on all 500kV substations, majority of the 220kV substations, and major power plants. With PMUs widely deployed in the system, one of JEPC's plans is to validate, improve, and update the network and generator parameters in both its EMS and PMS databases. Towards this end, JEPC recently developed an online PMU-assisted Power System Parameter Calibration System (PSPCS).

On one hand, the promise of synchrophasor measurements in system/parameter identification has been demonstrated in many previous work [1]-[6]. In [1]-[2], the authors presents methods of PMU-based transmission line (TL) parameter identification. Paper [3] discusses using PMU data for transformer parameter identification/calibration. Authors of [4] present the idea of PMU-based fault location. Papers [5]-[6] discuss PMU-based generator parameter identification. It is noted that these applications generally require PMU measurements to be very accurate.

On the other hand, although PMU data are expected to be highly accurate, the potential accuracy and reliability are not always achieved in actual field installation due to various causes [7]-[8]. JEPC has observed various types of data quality issues in PMU measurements under different occasions. To ensure accurate, reliable and consistent PMU data, there are pressing needs to calibrate PMU measurements to fulfill the claimed performance.

Considering both aspects, JEPC developed PSPCS to track TL/generator parameters and compare them against references obtained from EMS/PMS databases. The intent is not to determine which data source is more accurate, but rather detect and locate the discrepancies so that improvement can be made. Once a discrepancy is identified, a flag will be raised and further investigation will be initiated to figure out its cause(s), whether it is the reference values being out of date or the PMU data being problematic. In addition, from the management point of view, the EMS/PMS database are set up, maintained, and used by different teams/departments. In practice, due to the lack of communication, not all portion of the database stay updated. PSPCS provides a platform to bring all teams together towards the same goal to improve integrity of the system and accuracy of measurements. For example, unreported extension of a TL can be identified with help of PSPCS.

Further, the following two challenges need to be addressed: 1) there is currently no metric which can be used to evaluate the credibility of PMU based parameter estimation; 2) there is no practical method that can be used for online calibration of PMU measurements. A novel data mining framework is proposed to address these challenges. This paper presents the proposed methodologies, challenges and implementation issues identified during the development and deployment of PSPCS.

## II. MEASUREMENT MODELS

### A. Measurement Model for TL

As all PMUs in JEPC system output 3 phase quantities, a general 3-phase PI model is considered for TL in this work, as shown in Fig. 1. It could either be a nominal PI if the line is short or an equivalent PI if the line is longer. Variables in the figure are vectors of the 3-phase voltage/current phasors at sending and receiving end of the line and 3-by-3 impedance/admittance matrixes.

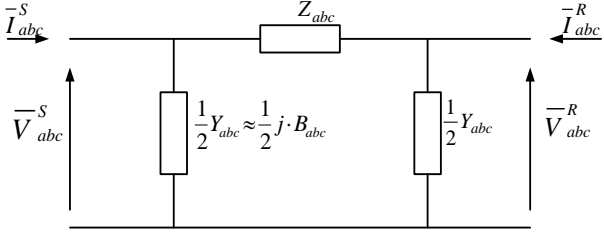

Fig. 1 3-phase transmission line PI model

The following equations are derived from nodal analysis:

$$\overline{V}^S_{abc} - \overline{V}^R_{abc} = Z_{abc} \cdot \overline{I}^S_{abc} - \frac{j}{2} \cdot Z_{abc} \cdot B_{abc} \cdot \overline{V}^S_{abc} \quad (1)$$

$$\overline{I}^S_{abc} + \overline{I}^R_{abc} = \frac{j \cdot B_{abc}}{2} \cdot (\overline{V}^S_{abc} + \overline{V}^R_{abc}) \quad (2)$$

Define matrix $Z_{abc}B_{abc}$ as $G_{abc}$ so that (1) becomes:

$$\overline{V}^S_{abc} - \overline{V}^R_{abc} = Z_{abc} \cdot \overline{I}^S_{abc} - \frac{j}{2} \cdot G_{abc} \cdot \overline{V}^S_{abc} \quad (3)$$

Matrix $G_{abc}$ is symmetrical as both $Z_{abc}$ and $B_{abc}$ are. Expanding (2) and (3) yields:

$$Z_a I^S_a + Z_{ab} I^S_b + Z_{ac} I^S_c \\ - j0.5 \cdot (G_a V^S_a + G_{ab} V^S_b + G_{ac} V^S_c) = V^S_a - V^R_a \quad (4)$$

$$Z_{ab} I^S_a + Z_b I^S_b + Z_{bc} I^S_c \\ - j0.5 \cdot (G_{ab} V^S_a + G_b V^S_b + G_{bc} V^S_c) = V^S_b - V^R_b \quad (5)$$

$$Z_{ac} I^S_a + Z_{bc} I^S_b + Z_c I^S_c \\ - j0.5 \cdot (G_{ac} V^S_a + G_{bc} V^S_b + G_c V^S_c) = V^S_c - V^R_c \quad (6)$$

$$2(I^S_a + I^R_a) - jB_a(V^S_a + V^R_a) - jB_{ab}(V^S_b + V^R_b) \\ - jB_{ac}(V^S_c + V^R_c) = 0 \quad (7)$$

$$2(I^S_b + I^R_b) - jB_{ab}(V^S_a + V^R_a) - jB_b(V^S_b + V^R_b) \\ - jB_{bc}(V^S_c + V^R_c) = 0 \quad (8)$$

$$2(I^S_b + I^R_b) - jB_{ac}(V^S_a + V^R_a) - jB_{bc}(V^S_b + V^R_b) \\ - jB_c(V^S_b + V^R_b) = 0 \quad (9)$$

Equations (4)-(9) are all complex and can be further expanded into 12 real one. With proper arrangement, these 12 equations can ultimately be put into the following standard form [9]:

$$Z = H \cdot x \quad (10)$$

where $H$ is a matrix and $Z$ is a vector, both of which are formulated from PMU measurements; $x$ is a vector made up of line impedances, which are to be estimated.

Therefore, with multiple sets of PMU measurements, the TL parameter identification problem can be formulated as a least-squares curve-fitting problem subject to inequality constraints:

$$\min_\beta \quad \frac{1}{2} \cdot \|H \cdot x - Z\|_2^2 \\ s.t. \quad lb_j \le x_j \le ub_j \quad j=1, 2, 3\dots \quad (11)$$

where $\|\bullet\|_2^2$ denotes square of the L2-norm of the corresponding vector, $j$ refers to the number of PMU measurements. Inequality constraints are set up based upon the EMS reference values. For example, it is reasonable to assume the to-be-estimated parameters lie in a certain error band (e.x., ±30%) of their corresponding references. Once $\beta$ is evaluated, TL sequence impedances can be obtained accordingly.

### B. Measurement Model for Generator

A synchronous generator model can be simplified as shown in Fig. 2 [10], where $H$ is the inertia constant, $K_R$ is the speed regulation constant, $K_D$ is the damping coefficient, $T$ is time constant of the turbine-governor, $w$ is the synchronous speed, $\delta$ is rotor angle, $P_e$ is electrical power, $P_m$ is mechanical power input and $P_{ref}$ is the power reference. Since objective of PSPCS is not to identify all parameters of a generator but to find out the discrepancy for improvement, this simplified model works well for such purpose.

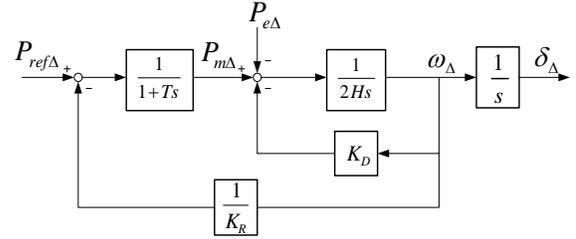

Fig. 2 A simplified turbine-generator model

The closed-loop transfer function can be identified as:

$$H(s) = \frac{\omega_\Delta}{P_{e\Delta}} = -\frac{1+Ts}{2HTs^2 + (2H+TK_D)s + (K_D + \frac{1}{K_R})} \quad (12)$$

The next step is to identify the discrete-time equivalent of the transform function given in (12). Assuming a sampling period (step size) of $h$, the following discretization equations can be used to approximate the first and second order derivatives of a continuous signal $x$ in time domain:

$$\frac{dx}{dt} \approx \frac{x(k) - x(k-1)}{h} \quad (13)$$

$$\frac{d^2x}{dt^2} = \frac{x(k+2) - 2x(k+1) + x(k)}{h^2} \quad (14)$$

where $k$ is the step number.

In time domain, (12) can be written as:

$$2HTw_\Delta^* + (2H + TK_D)w_\Delta' + (K_D + \frac{1}{K_R})w_\Delta = -P_{e\Delta} - TP_{e\Delta}' \quad (15)$$

Apply discretization rule (13)-(14) to (15), collect and rearrange items to obtain:

$$w_\Delta(k) = 2w_\Delta(k-1) - \frac{\left[2HT + 2Hh + ThK_D + h^2K_D + \frac{h^2}{K_R}\right]}{2HT} w_\Delta(k-2) + \frac{h(2H + TK_D)}{2HT} w_\Delta(k-3) - \frac{(h^2 + Th)}{2HT} P_{e\Delta}(k-2) + \frac{h}{2H} P_{e\Delta}(k-3) \quad (16)$$

This is essentially an autoregressive, moving average time series model (ARMA) which has standard form as:

$$y(k) = -a_1 y(k-1) - a_2 y(k-2) - a_3 y(k-3) + b_2 u(k-2) + b_3 u(k-3) + e(k) \quad (17)$$

Therefore, the standard coefficient identification process of ARMA model can be applied with multiple sets of PMU measurements [11]. Once the ARMA model solves, parameters of the generator can be evaluated:

$$\begin{cases} T = -\frac{b_3}{b_2 + b_3} \cdot h \\ H = \frac{h}{2b_3} \\ K_D = \frac{b_2 + b_3 - a_3 b_3}{b_3^2} \\ K_R = \frac{b_3^2(b_2 + b_3)}{b_3^2(-a_2 - a_3 + 1) - (b_2 + b_3)(b_2 + b_3 - a_3 b_3)} \end{cases} \quad (18)$$

### C. Error and Noise in PMU Measurements

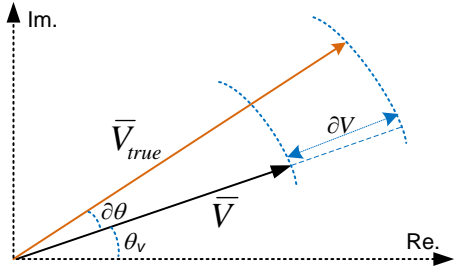

Fig. 3 Error in PMU measurement

In general, two types of errors can be identified in PMU measurements: random error and bias error. Random error, as the name suggests, is random (in either direction) in its nature and difficult to predict. Random error can be eliminated from measurements by statistical means. Systematic or bias error is reproducible inaccuracy that is consistently in the same direction. Bias error is much harder to estimate and remove. In PSPCS, both error types are considered and addressed in order to obtain meaningful results.

Fig. 3 shows a measured voltage phasor $\bar{V}$, its corresponding true value $\bar{V}_{true}$, and the associated error in terms of magnitude $\partial V$ and phase angle $\partial \theta$. The following relationship is derived:

$$\bar{V} = V \cdot e^{j\theta_V} \quad (19)$$

$$\bar{V}_{true} = (V + \partial V) \cdot e^{j(\theta_V + \partial \theta)} \quad (20)$$

where $V$ and $\theta_V$ are the magnitude and phase angle of phasor $\bar{V}$, respectively.

### III. CREDIBILITY MEASURE AND DATA CALIBRATION

#### A. Credibility Measure

Generally speaking, it is very difficult to determine the credibility of the calculated parameters obtained from PMU measurements, mainly because their true values are unknown and no comparison can be made. A metric which quantifies the credibility of the calculated parameters is of critical importance in practice for applications like such and so.

Bootstrapping is a re-sampling technique that can be used to estimate the properties of an estimator by sampling from an approximating distribution. In PSPCS, synchrophasor measurements collected during a certain time interval are used to conduct bootstrapping. By resampling with replacement, it is assumed that each sample of PMU measurements is independent and identically distributed. The measurement model for TL/generator is solved for each set of sampled PMU data to obtain one set of parameters. Conduct the sampling many times so that variance of each parameter can be evaluated. For example, statistics (probability density function) of a line reactance calculated from a bootstrapping procedure is visualized in Fig. 4. Ratio of the parameter's variance to the corresponding reference stored in the EMS/PMS database is used as the credibility metric. That is, the parameter obtained is credible if the corresponding metric is smaller than a pre-determined threshold:

$$\delta(x)/x_{ref} \leq \varepsilon \quad (21)$$

where $\delta(x)$ is variance of the calculated parameter, $x_{ref}$ is reference of $x$, and $\varepsilon$ is the predefined threshold.

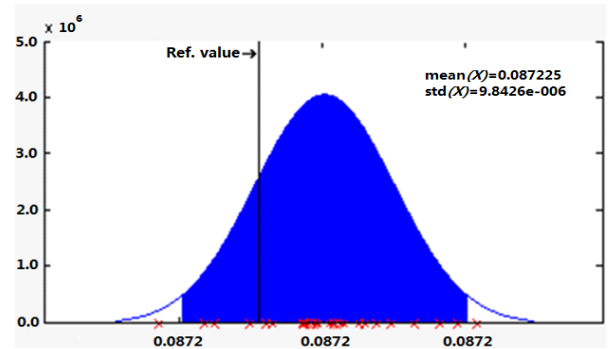

Fig. 4 Visualization of the credibility metric-statistics for line reactance

Flowchart for TL/generator screening within PSPCS is shown in Fig. 5. With bootstrapping and multiple PMU measurements, impact of random noise can be eliminated.

However, significant discrepancy can still be identified even if the calculated results are credible, which can be caused by two factors: 1) PMU measurements have bias errors; 2) EMS/PMS database is out of date. When discrepancy occurs, further investigation and data calibration are needed.

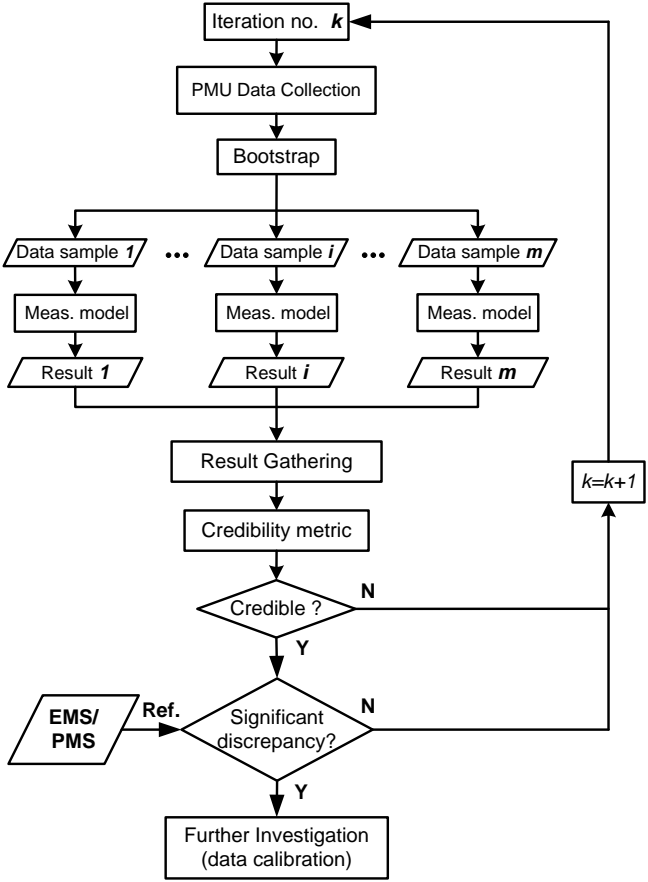

Fig. 5 PSPCS flowchart

### B. Data Calibration using DBSCAN

A novel data mining based synchrophasor measurement calibration framework is proposed during the development of PSPCS, which detects and corrects the overall systematic or bias error(s) introduced by PMU and its instrumentation channel. Major contribution of the proposed approach is that it does not require accurate knowledge of the system mathematical model/parameters. Due to space limitation, this subsection describes the proposed procedure for a TL and the procedure is similar for a generator.

To investigate the sensitivity of line parameters to bias errors in the both magnitude and phase angle of PMU measurements, partial derivatives need to be taken for the positive sequence components/portion of the nodal equations (1)-(2). Once the sensitivity factors are calculated, the bias error in PMU measurements and the errors in the TL reference values can be related using a linear relationship. Assuming multiple PMU measurements are collected and accurate TL parameters are known a priori, the bias errors in PMU measurements can be evaluated based on these sensitivity factors/linear relationship. The reality is that the accurate TL parameter is usually unknown. To address this challenge, data mining technique DBSCAN can be employed.

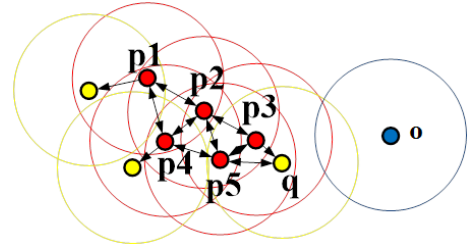

Fig. 6 Schematic diagram for DBSCAN

Basically, DBSCAN is a density-based spatial data clustering algorithm, the objective of which is to classify data points as core points, reachable points and outliers. As shown in Fig. 6, a core point $p$ contains a minimum number of points within the designed searching distance $\varepsilon$ (including $p$). A reachable point $q$ exists if there exist a path $p_1$->$p_2$->,…, $q$, and all the points on the path, except $q$, are core points. Point that is not reachable from any other point is outlier (point $o$). Core points and reachable points, as their names imply, are the meaningful data which form a cluster while outliers are not included in the cluster.

Although EMS values may be significantly wrong, our experience shows generally the error bands are well within 20%. Therefore, we define $\alpha$ as the error band multiplier for references obtained from the EMS database ($R_{EMS}$, $X_{EMS}$, and $B_{EMS}$). The following constraints can be considered:

$$\begin{cases}(1-\alpha)R_{EMS} \leq R \leq (1+\alpha)R_{EMS} \\ (1-\alpha)X_{EMS} \leq X \leq (1+\alpha)X_{EMS} \\ (1-\alpha)B_{EMS} \leq B \leq (1+\alpha)B_{EMS}\end{cases} \quad (22)$$

The corresponding feasibility region can be visualized as the cube shown in Fig. 7.

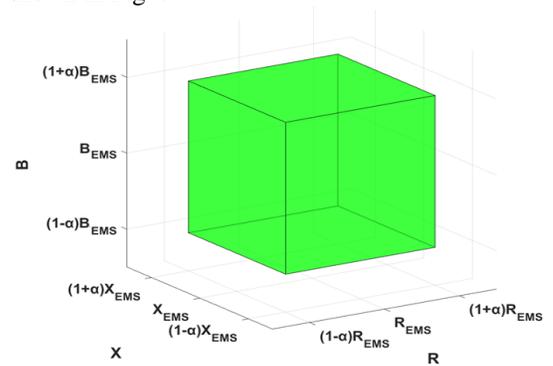

Fig. 7 Feasible region of the true TL parameters

The actual TL impedances should lie well within the feasible region shown in Fig. 7. However, only one point within this feasible region corresponds to the true TL parameters. The basic idea of the proposed approach is to 1) scan every point within the feasible region (a total of $M$ points); 2) evaluate the corresponding bias errors in the PMU measurements; 3) form sets of points with each set containing seven data made up by the bias errors in PMU measurements (a total of *M* sets). 4) apply DBSCAN to cluster all the *M* data sets to find out the one with least number of outliers (maximum number of core and reachable points) and minimum searching distance. Once this cluster is identified, the actual bias errors in all PMU channels and errors in line impedance references can be determined accordingly.

TABLE 1
RESULTS FOR DATA CALIBRATION

| Bias Error in | Ideal $\times 10^{-3}$ | Calculated $\times 10^{-3}$ | $R$ | $X$ | $B$ |
|---|---|---|---|---|---|
| $V_s$ | 0 | 0.00 | $0.98 \cdot R_{EMS}$ | $X_{EMS}$ | $B_{EMS}$ |
| $V_r$ | 0 | 0.00 | | | |
| $I_s$ | 10 | 10.00 | | | |
| $I_r$ | 0 | 0.00 | | | |
| $\theta_{Vs}$ | 0 | 0.00 | | | |
| $\theta_{Vr}$ | 0 | 0.00 | | | |
| $\theta_{Is}$ | 0 | 0.00 | | | |
| $\theta_{Ir}$ | 0 | 0.00 | | | |

## IV. NUMERICAL EXAMPLE

A numerical example is provided in this section to illustrate the procedure of PMU data calibration. A simulation model with a TL, two PMUs, and time-varying load is built in Matlab/Simulink. In the experiment, it is assumed that the series reactance and shunt susceptance of the TL obtained from the EMS database are both accurate ($X_{EMS}=X_{true}$, $B_{EMS}=B_{true}$) and there is an error in the line resistance value ($R_{EMS}=0.98R_{true}$). A 0.01 p.u. bias error in the magnitude of the sending-end current measurements is considered (added to the PMU measurements).

Using the proposed data calibration approach, set $\alpha$ =10%, and implement DBSCAN algorithm by varying the reference value of $R$ from 0.9 $R_{EMS}$ to 1.1 $R_{EMS}$. The bias error for each PMU channel is continuously evaluated for every scanned $R$ value and relationship between each scanned $R$ value and the corresponding bias errors in all PMU channels are shown in Fig. 8.

Results of the experiment is summarized in Table 1. Fig. 8 shows the core points and the outlier, based on which the bias error in the PMU measurements is identified to be in the magnitude of the sending-end current phasors. The actual line resistance is found to be 0.98 times of the given reference value, as expected. Therefore, both bias error in PMU measurements and the true impedance value of the TL can be identified through the proposed data mining based approach.

## V. CONCLUSION

PMU has the potential to improve the accuracy of power system models and parameters. More accurate parameter means better power system modeling, faster and more accurate fault location as well as more economic system operation. Jiangsu Electric Power Company recently developed an online power system parameter calibration system (PSPCS), which conducts real-time screening of the EMS and PMS database to identify discrepancies between the calculated parameters and the reference values. Once significant discrepancy is observed, further investigation including data calibration will be conducted to identify the cause of the problem and improvement will be made accordingly.

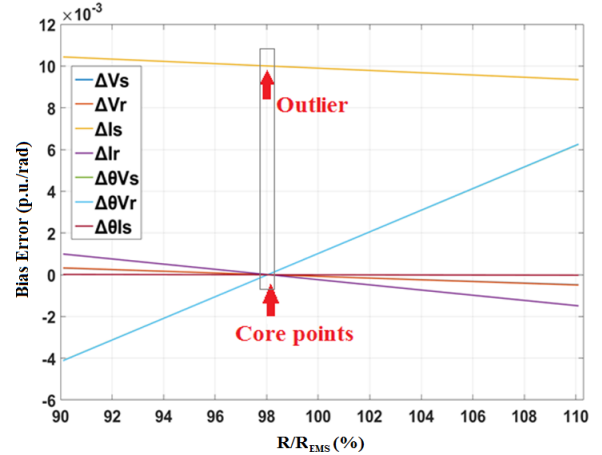

Fig. 8 Results from DBSCAN

## VI. ACKNOWLEDGEMENT

This work is funded by SGCC Science and Technology Program under contract no. 5455HJ160007.